\documentclass[runningheads]{llncs}
\usepackage[rgb,x11names]{xcolor}
\usepackage{amssymb}
\usepackage{amsmath}
\usepackage{subfig}
\usepackage{hyperref}
\hypersetup{colorlinks=true,citecolor=blue}
\usepackage{graphicx}
\usepackage[]{algorithm2e}
\usepackage{mathrsfs}
\usepackage{bbding}
\usepackage{rotating}
\usepackage{multirow}
\usepackage{marginnote}
\usepackage{overpic}
\usepackage{colortbl}
\usepackage[labelfont=bf]{caption}
\usepackage{bbding}

\newcommand{\secref}[1]{$\S$ \ref{#1}}

\newcommand{\figref}[1]{Fig. \ref{#1}}
\newcommand{\tabref}[1]{Tab. \ref{#1}}

\definecolor{mygray}{gray}{.92}
\def\etal{{\em et al.}}
\def\ie{\emph{i.e.}}
\def\eg{\emph{e.g.}}
\def\etc{\emph{etc}}
\def\etal{{\em et al.~}}

\def\ourmodel{\textit{PraNet}}

\makeatletter
\newcommand\figcaption{\def\@captype{figure}\caption}
\newcommand\tabcaption{\def\@captype{table}\caption}
\makeatother

\usepackage{graphicx}
\graphicspath{{./Imgs/}}
\DeclareGraphicsExtensions{.jpg,.pdf,.png}

\begin{document}
%
%\title{CRANet: Cascade Reverse Attention \\for polyp segmentation}
%\title{PraNet: parallel Reverse Attention \\for Polyp Segmentation}
%\title{PraNet for Fast and Accurate \\Polyp Segmentation}
%\title{Pra: Parallel Reverse Attention Architecture \\for Medical Image Segmentation}
\title{PraNet: Parallel Reverse Attention Network \\for Polyp Segmentation}

%
%\titlerunning{Abbreviated paper title}
% If the paper title is too long for the running head, you can set
% an abbreviated paper title here
%
%\author{Deng-Ping Fan\inst{1}\orcidID{0000-1111-2222-3333} \and
%Second Author\inst{2,3}\orcidID{1111-2222-3333-4444} \and
%Third Author\inst{3}\orcidID{2222n/a3333-4444-5555}}

\author{Deng-Ping Fan\inst{1} % index{Deng-Ping, Fan}
\and
Ge-Peng Ji\inst{2} % index{Ge-Peng, Ji}
\and
Tao Zhou\inst{1}   % index{Tao, Zhou}
\and 
Geng Chen\inst{1}  % index{Geng, Chen}
\and  \\
Huazhu Fu\inst{1}  % index{Huazhu, Fu}
\Envelope \and
Jianbing Shen\inst{1} % index{Jianbing, Shen}
\Envelope \and
Ling Shao          % index{Ling, Shao}
\inst{3,1}
}
%

%\author{Anonymous}
%\authorrunning{Paper ID \#$1031$}
\institute{$^1$ Inception Institute of Artificial Intelligence, Abu Dhabi, UAE.\\
$^2$ School of Computer Science, Wuhan University, Hubei, China. \\
$^3$ Mohamed bin Zayed University of Artificial Intelligence, Abu Dhabi, UAE. \\
{\tt \{huazhu.fu, jianbing.shen\}@inceptioniai.org}\\
{\tt \href{https://github.com/DengPingFan/PraNet}{https://github.com/DengPingFan/PraNet}}
}

\authorrunning{D.-P. Fan et al.}

\titlerunning{PraNet: Parallel Reverse Attention Network for Polyp Segmentation}

% First names are abbreviated in the running head.
% If there are more than two authors, 'et al.' is used.
%
%\institute{Princeton University, Princeton NJ 08544, USA \and
%Springer Heidelberg, Tiergartenstr. 17, 69121 Heidelberg, Germany
%\email{lncs@springer.com}\\
%\url{http://www.springer.com/gp/computer-science/lncs} \and
%ABC Institute, Rupert-Karls-University Heidelberg, Heidelberg, Germany\\
%\email{\{abc,lncs\}@uni-heidelberg.de}}

\maketitle              % typeset the header of the contribution
\begin{abstract}
Colonoscopy is an effective technique for detecting colorectal polyps, which are highly related to colorectal cancer.
In clinical practice, segmenting polyps from colonoscopy images is of great importance since it provides valuable information for diagnosis and surgery.
However, accurate polyp segmentation is a challenging task, for two major reasons: (i) the same type of polyps has a diversity of size, color and texture; and (ii) the boundary between a polyp and its surrounding mucosa is not sharp. 
To address these challenges, we propose a parallel reverse attention network (\ourmodel) for accurate polyp segmentation in colonoscopy images.
Specifically, we first aggregate the features in high-level layers using a parallel partial decoder (PPD). Based on the combined feature, we then generate a global map as the initial \textit{\textbf{guidance area}} for the following components.
In addition, we mine the \textit{\textbf{boundary cues}} using the reverse attention (RA) module, which is able to establish the relationship between areas and boundary cues.
Thanks to the recurrent cooperation mechanism between areas and boundaries, our \ourmodel~is capable of calibrating some misaligned predictions, improving the segmentation accuracy.
%PraNet aggregates high-level features via a parallel partial decoder and a reverse attention module.
Quantitative and qualitative evaluations on five challenging datasets across six metrics show that our \ourmodel~improves the segmentation accuracy significantly, and presents a number of advantages in terms of generalizability, and real-time segmentation efficiency (\textbf{$\sim$50fps}). 
% {\em All the codes and results will be released.} 
%upon the publication of this work.
%, and training economy. %(\textbf{$\sim$0.5 hours}). % at \textcolor{red}{*****} %\url{http://hidden.for.anonymity}.
%Codes will be made publicly available soon.
% ($>$5\% imp. in Dice $vs.$ the cutting-edge model: ResUNet++)

%\and Reverse attention \and Partial decoder \and parallel connection} % \and PraNet.}%\vspace{-0.35cm}
\end{abstract}
\section{Introduction}
Colorectal cancer (CRC) is the third most common type of cancer around the world \cite{silva2014toward}. %after lung cancer and breast cancer.
Therefore, preventing CRC by screening tests and removal of preneoplastic lesions (colorectal adenomas) is very critical and has become a worldwide public health priority.
Colonoscopy is an effective technique for CRC screening and prevention since it can provide the location and appearance information of colorectal polyps, enabling doctors to remove these before they develop into CRC. A number of studies have shown that early colonoscopy has contributed to a 30\% decline in the incidence of CRC \cite{haggar2009colorectal}.
Thus, in a clinical setting, accurate polyp segmentation is of great importance. It is a challenging task, however, due to two major reasons. First, the polyps often vary in appearance, \eg, size, color and texture, even if they are of the same type. Second, in colonoscopy images, the boundary between a polyp and its surrounding mucosa is usually blurred and lacks the intense contrast required for segmentation approaches. These issues result in the inaccurate segmentation of polyps, and sometimes even cause the missing detection of polyps.
Therefore, an automatic and accurate polyp segmentation approach capable of detecting all possible polyps at an early stage is of great significance in the prevention of CRC \cite{jia2019wireless}.

Among the various polyp segmentation methods, 
%learning-based ones have shown particular good performance.
the early learning-based methods rely on extracted hand-crafted features \cite{mamonov2014automated,tajbakhsh2015automated}, such as color, texture, shape, appearance, or a combination of these features. These methods are usually trained a classifier to distinguish a polyp from its surroundings.  
However, these models often suffer from a high miss-detection rate. The main reason is that the representation capability of hand-crafted features is quite limited when it comes to dealing with the high intra-class variations of polyps and low inter-class variations between polyps and hard mimics \cite{yu2016integrating}. 
Recently, numerous deep learning based methods have been developed for polyp segmentation~\cite{yu2016integrating,zhang2018polyp}. Although progress has been made by these methods, they only detect polyps using a bounding boxes, thus failing to locate accurate boundaries of polyps. To address this issue, Brandao \etal \cite{brandao2017fully} employed an FCN with a pre-trained model to identify and segment polyps. Akbari \etal \cite{akbari2018polyp} utilized a modified version of FCN to improve the accuracy of polyp segmentation.
Inspired by the success of the U-Net~\cite{ronneberger2015u} applied in biomedical image segmentation, U-Net++~\cite{zhou2018unetplus} and ResUNet++~\cite{jha2019resunetplus} were employed for polyp segmentation and obtained promising performance. These methods focus on segmenting the whole area of the polyp, but they ignore the area-boundary constraint, which is very critical for enhancing the segmentation performance. To this end, Psi-Net~\cite{murugesan2019psi} utilized area and boundary information simultaneously in polyp segmentation, but the relationship between the area and boundary was not fully captured. Besides, Fang \etal \cite{fang2019selective} proposed a three-step selective feature aggregation network with area and boundary constraints for polyp segmentation. This method \textit{explicitly} considers the dependency between areas and boundaries and obtains good results with additional edge supervision; however, it is time-consuming ($>$20 hours) and easily corrupted with over-fitting.% (\tabref{tab:Generalizability}).

%it can predict them simultaneously without iterative modeling their dependency during the training stage. As a results, it could fail to segment the non-sharp boundaries. 

%Tajbakhsh \etal \cite{tajbakhsh2015automatic} employed convolutional neural networks for learning discriminative spatial and temporal features. Yu \etal \cite{yu2016integrating} proposed an effective 3D-fully convolutional network (FCN) to learn spatio-temporal feature representations for polyp detection from colonoscopy videos. Zhang \etal \cite{zhang2018polyp} proposed to use regression-based deep CNN methods with residual learning for polyp detection. 

%Brandao \etal \cite{brandao2017fully} employed the FCN with a pre-trained VGG model to identify and segment polyps. Akbari \etal \cite{akbari2018polyp} proposed a polyp segmentation method based on convolutional neural network and Otsu thresholding. Jha \etal \cite{jha2019resunetplus} proposed an improved ResUNet architecture (ResUNet++) for colonoscopic image segmentation. Fang \etal \cite{fang2019selective} proposed a novel selective feature aggregation network with the area and boundary constraints for polyp segmentation.
%Although these deep learning based methods have achieved promising advancements,
%they share a common drawbacks. First, most of them overlook the importance of area and  
%it is still challenging to accurately locate the boundary of polyps from the surrounding mucosa.

In this paper, we propose a novel deep neural network, called \textbf{P}arallel \textbf{R}everse \textbf{A}ttention \textbf{Net}work (\textbf{\ourmodel}), for the polyp segmentation task.
Our motivation stems from the fact that, during polyp annotation, clinicians first roughly locate a polyp and then accurately extract its silhouette mask according to the local features. We therefore argue that the area and boundary are two key characteristics that distinguish normal tissues and polyps.
%This idea is also supported by the latest work in~\cite{fang2019selective}, which \textit{explicitly} constructs a shared encoder and two tightly constrained decoder focusing on polyp areas and boundaries, respectively.
%
Different from \cite{fang2019selective}, we first predict coarse areas and then \textit{implicitly} model the boundaries by means of reverse attention.
There are three advantages to this strategy, including better learning ability, improved generalization capability, and higher training efficiency. Please refer to our experiments (\secref{sec:Experiments}) for more details.
In a nutshell, our contributions are threefold. 
(1) We present a novel deep neural network for real-time and accurate polyp segmentation. By aggregating features in high-level layers using a parallel partial decoder (PPD), the combined feature takes contextual information and generates a global map as the initial \textit{\textbf{guidance area}} for the subsequent steps. To further mine the \textit{\textbf{boundary cues}}, we leverage a set of recurrent reverse attention (RA) modules to establish the relationship between areas and boundary cues. Due to this  recurrent cooperation mechanism between areas and boundaries, our model is capable of calibrating some misaligned predictions. 
(2) We introduce several novel evaluation metrics for polyp segmentation and present a comprehensive benchmark for existing SOTA models that are publicly available.
(3) Extensive experiments demonstrate that the proposed \ourmodel~outperforms most cutting-edge models and advances the SOTAs by a large margin, on five challenging datasets, with real-time inference and shorter training time. %The codes and benchmark results have been released at 

%The proposed model consistently outperforms all state-of-the-arts on the new challenging
%Although several methods have been developed for this task, there still exist several issues. \emph{First}, conventional prediction approaches rely on radiomic

%Training a medical deep learning model usually takes tens of hours, or even days. In order to get better detection accuracy, often the parameters of the model will be improved a lot. How to develop a fast and good depth model is of great significance to the development of the field

\begin{figure*}[t!]
	\centering
    %\small
	\begin{overpic}[width=\textwidth]{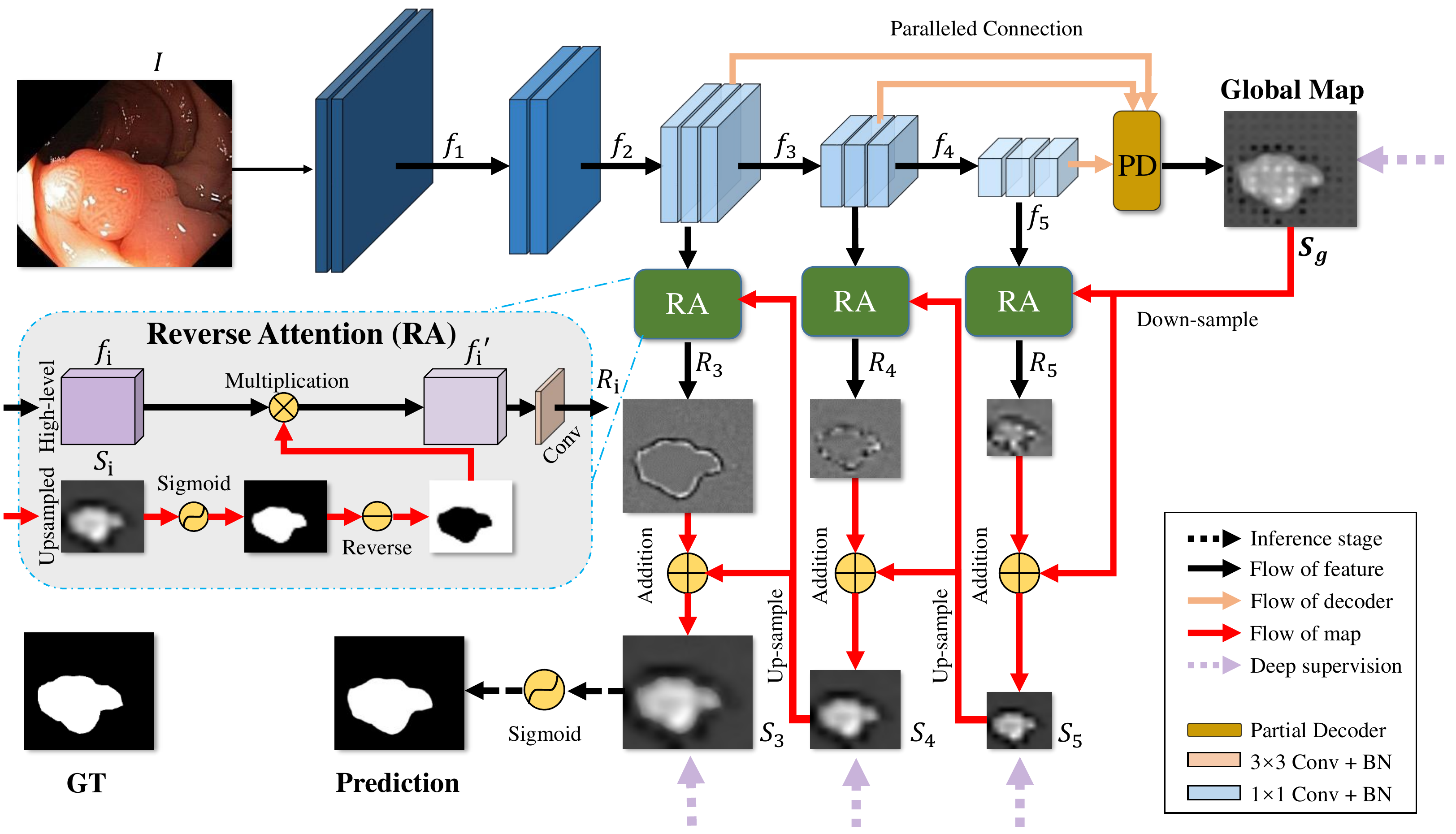}
    \end{overpic}
	\caption{Overview of the proposed \ourmodel,~which consists of three reverse attention modules with a parallel partial decoder connection. See \secref{sec:Methods} for details.}
    \label{fig:Framework}
\end{figure*}

\section{Method}\label{sec:Methods}
\figref{fig:Framework} shows our \ourmodel, which utilizes a  parallel partial decoder to generate the high-level semantic global map and a set of reverse attention modules for accurate polyp segmentation from the colonoscopy images. Each component will be elaborated as follows. %of the proposed model will be elaborated next.

\subsection{Feature Aggregating via Parallel Partial Decoder}\label{sec:PPD}
Current popular medical image segmentation networks usually rely on a U-Net~\cite{ronneberger2015u} or a U-Net like network (\eg, U-Net++~\cite{zhou2018unetplus}, ResUNet~\cite{zhang2018road}, \etc). 
These models are essentially encoder-decoder frameworks, which typically aggregate \textit{all} multi-level features extracted from CNNs. As demonstrated by Wu~\etal\cite{wu2019cascaded}, compared with high-level features, low-level features demand more computational resources due to their larger spatial resolutions, but contribute less to performance.
Motivated by this observation, we propose to aggregate high-level features with a \textbf{\textit{parallel partial decoder}} component. More specifically, for an input polyp image $I$ with size $h\times w$, five levels of features $\{\textbf{f}_i, i = 1,...,5 \}$ with resolution $[h/2^{k-1}, w/2^{k-1}]$ can be extracted from Res2Net-based~\cite{pami20Res2net} backbone network.
%(\eg, VGGNet~\cite{simonyan2015very}, ResNet~\cite{he2016deep}, or Res2Net~\cite{pami20Res2net}, \etc). 
Then, we divide $\textbf{f}_i$ features into low-level features $\{\textbf{f}_i,i = 1,2\}$ and high-level features $\{\textbf{f}_i, i = 3,4,5\}$. We introduce the partial decoder $p_d (\cdot)$~\cite{wu2019cascaded}, a new SOTA decoder component, to aggregate the high-level features with a paralleled connection. The partial decoder feature is computed by $\textbf{PD} = p_d(f_3,f_4, f_5)$, and we can obtain a global map $\textbf{S}_g$.

\subsection{Reverse Attention Module}
In a clinical setting, doctors first roughly locate the polyp region, and then carefully inspect local tissues to accurately label the polyp.
As discussed in \secref{sec:PPD}, our global map $\textbf{S}_g$ is derived from the deepest CNN layer, which can only capture a relatively rough location of the polyp tissues, without structural details (see \figref{fig:Framework}).
To address this issue, we propose a principle strategy to progressively mine discriminative polyp regions through an erasing foreground object manner~\cite{wei2017object,chen2018reverse}. Instead of aggregating features from all levels like in~\cite{chen2018reverse,Gu2019,ETNet2019,AGNet}, we propose to adaptively learn the \textit{\textbf{reverse attention}} in three parallel high-level features. In other words, our architecture can sequentially mine complementary 
regions and details by erasing the existing estimated polyp regions from high-level side-output features, where the existing estimation is up-sampled from the deeper layer.

Specifically, we obtain the output reverse attention features $R_i$ by multiplying  (element-wise $\odot$) the high-level side-output feature $\{f_i, i=3,4,5\}$ by a reverse attention weight $A_i$, as below:
%\vspace{-0.4cm}
\begin{equation} \label{RA}
R_i = f_i \odot A_i.
\end{equation}
The reverse attention weight $A_i$ is de-facto for salient object detection in the computer vision community~\cite{chen2018reverse,zhang2020bilateral}, and can be formulated as:
\begin{equation} \label{WeightAttention}
A_i = \circleddash(\sigma(\mathcal{P}(S_{i+1}))),
\end{equation}
where $\mathcal{P}(\cdot)$ denotes an up-sampling operation, $\sigma(\cdot)$ is the Sigmoid function, and $\circleddash(\cdot)$ is a reverse operation subtracting the input from matrix $\mathbf{E}$, in which all the elements are $1$. 
\figref{fig:Framework} (RA) shows the details of this process.
It is worth noting that the erasing strategy driven by reverse attention can eventually refine the imprecise and coarse estimation into an accurate and complete prediction map. 

\subsection{Learning Process and Implementation Details.}

\noindent\textbf{Loss Function.} Our loss function is defined as $\mathcal{L} = \mathcal{L}_{IoU}^w + \mathcal{L}_{BCE}^w$, where $\mathcal{L}_{IoU}^w$ and $\mathcal{L}_{BCE}^w$ represent the weighted IoU loss and binary cross entropy (BCE) loss for the global restriction and local (pixel-level) restriction. Different from the standard IoU loss, which has been widely adopted in segmentation tasks, the weighted IoU loss increases the weights of hard pixels 
to highlight their importance. In addition, compared with the standard BCE loss, $\mathcal{L}_{BCE}^w$ pays more attention to hard pixels rather than assigning all pixels equal weights. The definitions of these losses are the same as in~\cite{qin2019basnet,wei2019f3net} and their effectiveness has been validated in the field of salient object detection. Here, we adopt deep supervision for the three side-outputs (\ie, $S_3$, $S_4$, and $S_4$) and the global map $S_g$. Each map is up-sampled (\eg, $S_3^{up}$) to the same size as the ground-truth map $G$. Thus the total loss for the proposed \ourmodel~can be formulated as:
$\mathcal{L}_{total} = \mathcal{L} (G, S_g^{up}) + \sum_{i=3}^{i=5} \mathcal{L} (G, S_i^{up})$. \\

\noindent\textbf{Implementation Details.}
%Firstly, we split the Kvasir~\cite{jha2020kvasir} dataset into training/testing part (90\%/10\%) like~\cite{jha2019resunetplus}. 
%We use Res2Net~\cite{pami20Res2net} as our backbone and 
We implement our model in PyTorch, which is accelerated by an NVIDIA TITAN RTX GPU. All the inputs are uniformly resized to $352 \times 352$ and employ a multi-scale training strategy $\{0.75, 1, 1.25\}$ rather than data augmentation. We employ the Adam optimization algorithm to optimize the overall parameters with a learning rate of $1e-4$. The whole network is trained in an end-to-end manner, which takes 32 minutes to converge  over 20 epochs with a batch size of 16. Our final prediction map $S_p$ is generated by $S_3$ after a sigmoid operation.
% With limit computation source, if we want to train a deep-based medical architecture requires long time (\eg, 1$\sim$)

%Time;
%Speed
%low computational 
%Train fast

\section{Experiments}\label{sec:Experiments}

\subsection{Experiments on Polyp Segmentation}\label{sec:CmpSOTAS}
In this section, we compare our \ourmodel~with existing methods in terms of learning ability, generalization capability, complexity, and qualitative results. \\

\noindent\textbf{Datasets and Baselines.} Experiments are conducted on five polyp segmentation datasets: ETIS~\cite{silva2014toward},
CVC-ClinicDB/CVC-612~\cite{bernal2015wm}, CVC-ColonDB~\cite{tajbakhsh2015automated}, 
EndoScene~\cite{vazquez2017benchmark}, and Kvasir~\cite{jha2020kvasir}. 
The first four are standard benchmarks, and the last one is the largest-scale challenging dataset, recently released. We compare our \ourmodel~with four SOTA medical image segmentation methods: U-Net~\cite{ronneberger2015u}, U-Net++~\cite{zhou2018unetplus}, ResUNet-mod~\cite{zhang2018road}, and ResUNet++~\cite{jha2019resunetplus}. We also report the cutting edge polyp segmentation model, \ie, SFA~\cite{fang2019selective}. 
The segment results of SFA are generated by the released code with default settings. \\

\noindent\textbf{Training Settings and Metrics.} Unless otherwise noted, we follow the same training settings as in~\cite{jha2019resunetplus}, \ie, the images from Kvasir, and CVC-ClinicDB are randomly split into 80\% for training, 10\% for validation, and 10\% for testing.  
We employ two metrics (\ie, mean Dice and mean IoU) for quantitative evaluation, similar to~\cite{jha2019resunetplus,jha2020kvasir}. To provide deeper insight into the model performance, we further introduce four other metrics which are widely used in the field of object detection~\cite{fan2020camouflaged,Zhang2020UCNet,fan2018salient,fu2020jl,zhao2019contrast,zhao2019egnet}. The weighted Dice metric $F_\beta^w$ is used to amend the ``Equal-importance flaw'' in Dice. The MAE metric is utilized to evaluate the pixel-level accuracy. To evaluate pixel-level and global-level similarity, we adopt the recently released enhanced-alignment metric $E_\phi^{max}$~\cite{Fan2018Enhanced}. Since $F_\beta^w$ and MAE are based on a pixel-wise evaluation system and ignore structural similarities, $S_{\alpha}$~\cite{fan2017structure} is adopted to assess the similarity between predictions and ground-truths. The evaluation toolbox is available at {\tt \href{https://github.com/DengPingFan/PraNet}{https://github.com/DengPingFan/PraNet}}.
\\ 

%(\ie, $F_\beta^w$, $S_{\alpha}$, $E_\phi^{max}$, and MAE), 
 
%same evaluation metrics as~\cite{jha2019resunetplus,jha2020kvasir}, \ie, 1) Sensitivity/Recall, 2) Precision, 3) mean intersection over union (IoU), 4) Dice coefficient. 
%To provide a more comprehensive understanding of the performance, we also introduce four important metrics adopted in ~\cite{Zhang2020UCNet} (\ie, $F_\beta$, $F_\beta^w$, $S_{\alpha}$, and $E_\phi^{max}$) for foreground and semantic segmentation tasks. \\
%$F_\beta$-score~\cite{Zhang2020UCNet,Fu2020JLDCF}, $F_\beta^w$-score~\cite{margolin2014evaluate}, $S_{\alpha}$-score~\cite{fan2017structure}, and $E_\phi^{max}$-score~\cite{Fan2018Enhanced}) for foreground and semantic segmentation tasks. \\

\begin{table*}[t!]
  \centering
  \scriptsize
  \renewcommand{\arraystretch}{1.1}
  \setlength\tabcolsep{5pt}
  \caption{Quantitative results on Kvasir~\cite{jha2020kvasir} and  CVC-612~\cite{bernal2015wm} datasets.
  `n/a' denotes that the results are not available. `$\dagger$' represents evaluation scores from~\cite{jha2019resunetplus}.
  %Bold text presents the best score. 
  %We also train and test existing cutting edge model SFA~\cite{fang2019selective} from scratch under the same setting as ours for a fair comparison.
  %(\eg, U-Net and U-Net++) under the same setting for a fair comparison.
  }\label{tab:Kvasir-CVC612}
  \begin{tabular}{cr||cccccccc}
  \hline
  % \multicolumn{1}{l|}{Methods} &\multicolumn{3}{c|}{\emph{MCL}}\\
  \rowcolor{mygray}
   &Methods & mean Dice & mean IoU  &  $F_\beta^w$ & $S_{\alpha}$&$E_\phi^{max}$ & MAE\\
  \hline
  %\multicolumn{7}{c}{Current state-of-the-art polyp segmentation results} \\
  %\hline
  \multirow{6}{*}{\begin{sideways}Kvasir\end{sideways}} & 
  %U-Net$^\dagger$~\cite{ronneberger2015u}  & 0.715 & 0.433 & n/a & n/a & n/a & n/a\\
  U-Net (MICCAI'15)~\cite{ronneberger2015u} & 0.818 & 0.746 & 0.794 & 0.858 & 0.893 & 0.055 \\
  &U-Net++ (TMI'19)~\cite{zhou2018unetplus} & 0.821  & 0.743 & 0.808 & 0.862 & 0.910 & 0.048 \\
  %ResUNet$^\dagger$~\cite{zhang2018road}  & 0.504 & 0.729 & 0.608 & 0.436 & 0.514 & n/a & n/a & n/a & n/a\\
  & ResUNet-mod$^\dagger$~\cite{zhang2018road} & 0.791 &  n/a  & n/a & n/a & n/a & n/a \\
  & ResUNet++$^\dagger$~\cite{jha2019resunetplus} & 0.813 & 0.793 & n/a & n/a & n/a & n/a\\
  & SFA (MICCAI'19)~\cite{fang2019selective} & 0.723 & 0.611 & 0.670 & 0.782 & 0.849 & 0.075\\
  & \textbf{\ourmodel~(Ours)} & \textbf{0.898} & \textbf{0.840} & \textbf{0.885} & \textbf{0.915} & \textbf{0.948} & \textbf{0.030}\\
  %\hline
  %\multicolumn{7}{c}{Classical medical image segmentation baselines trained from scratch} \\
  %\hline
  %U-Net~\cite{ronneberger2015u} & 0.832 & 0.712 & 0.793 & 0.857 & 0.892 & 0.055 \\
  %U-Net++~\cite{zhou2018unetplus} & 0.836  & 0.719 & 0.808 & 0.862 & 0.911 & 0.048 \\
  %SFA-MICCAI'19~\cite{fang2019selective} & 0.718 & 0.561 & 0.670 & 0.782 & 0.849 & 0.075\\
  %CPD-CVPR'19~\cite{wu2019cascaded} & 0.900 &  0.889& 0.874 & 0.782 & 0.874& 0.856 & 0.894 & 0.931 & 0.039 \\
  %SCRNet-ICCV'19~\cite{wu2019stacked} & 0.911 & 0.874 & 0.873 & 0.804 & 0.873 & 0.851 & 0.905 & 0.891 & 0.035 \\
  %SINet-CVPR'20~\cite{fan2020Camouflage} & 0.915 & 0.851 & 0.861 & 0.791 & 0.861 & 0.809 & 0.902 & 0.938 & 0.036\\
  %\textbf{\ourmodel~(Ours)} & \textbf{0.896} & \textbf{0.824} & \textbf{0.885} & \textbf{0.915} & \textbf{0.948} & \textbf{0.030}\\
  \hline
  \hline
  \multirow{6}{*}{\begin{sideways}CVC-612\end{sideways}} & 
  %U-Net$^\dagger$~\cite{ronneberger2015u} & 0.642 & 0.471  & n/a & n/a & n/a & n/a \\
  %ResUNet$^\dagger$~\cite{zhang2018road}  & 0.578 & 0.561 & 0.627 & 0.457 & 0.451 & n/a & n/a & n/a & n/a\\
  U-Net (MICCAI'15)~\cite{ronneberger2015u} & 0.823 & 0.755 & 0.811 & 0.889 & 0.954 & 0.019 \\
  &U-Net++ (TMI'19)~\cite{zhou2018unetplus} & 0.794  & 0.729 & 0.785 & 0.873 & 0.931 & 0.022\\
  &ResUNet-mod$^\dagger$~\cite{zhang2018road} & 0.779 & n/a & n/a & n/a & n/a & n/a \\
  &ResUNet++$^\dagger$~\cite{jha2019resunetplus} & 0.796  & 0.796 & n/a & n/a & n/a & n/a\\
  &SFA (MICCAI'19)~\cite{fang2019selective} & 0.700 & 0.607 & 0.647 & 0.793 & 0.885 & 0.042\\
  &\textbf{\ourmodel~(Ours)} & \textbf{0.899} & \textbf{0.849} & \textbf{0.896} & \textbf{0.936} & \textbf{0.979}  & \textbf{0.009}\\
  \hline
  \end{tabular}
\end{table*}

\noindent\textbf{Learning Ability.} In this section, we conduct two experiments to validate our model's learning ability on two \textit{seen} datasets, \ie, Kvasir and CVC-612.
%Since our model have \textit{seen} (trained) on Kvasir and CVC-612 datasets, thus we need to conduct two experiments to test the learning ability. 
%$\bullet$~
\textit{Kvasir} is a recently released challenging dataset that contains 1,000 images selected from a sub-class (polyp class) of the Kvasir dataset~\cite{pogorelov2017kvasir}. 
\textit{CVC-ClinicDB}, also called \textit{CVC-612}, includes 612 open-access images from 31 colonoscopy clips.
As shown in \tabref{tab:Kvasir-CVC612}, our \ourmodel~outperforms all SOTAs by a large margin (mean Dice: about $>7\%$), across both datasets, in all metrics. 
This suggests that our model has a strong learning ability to effectively segment polyps. \\

\noindent\textbf{Generalization Capability.} We conduct three experiments to test the model's generalizability. The three \textit{unseen} datasets have their own challenging situations and properties. \textit{CVC-ColonDB} is a small-scale database which contains 380 images from 15 short colonoscopy sequences. All images are used as our testing set.
%of annotated video sequences of colonoscopy video. It contains 15 short colonoscopy sequences, coming from 15 different studies.
\textit{ETIS} is an early established dataset which has 196 polyp images for early diagnosis of colorectal cancer.
\textit{EndoScene} is a combination of CVC-612 and CVC300. We follow Fang~\etal~\cite{fang2019selective} and split it into training, validation, and testing subsets. We only use the testing set of EndoScene-CVC300 in this experiment, since part of CVC-612 may be seen in the training stage. 
\ourmodel~ again outperforms existing classical medical segmentation baselines (\ie, U-Net, U-Net++), as well as SFA, with significant improvements (see \tabref{tab:Generalizability})
on all three unseen datasets. One notable finding is that SFA drops dramatically on these unseen datasets, partially demonstrating that the model generalizability is poor. \\

%listed in~\cite{jha2019resunetplus}. 
%To the best of our knowledge, there is no deep-based polyp segmentation models are publicly available. 
%We, therefore, introduce three representative SOTA models for different tasks (\ie,  medical image segmentation~\cite{zhou2018unetplus}, camouflage object segmentation~\cite{fan2020Camouflage}, and salient object segmentation~\cite{wu2019stacked}). 

\begin{table*}[t!]
  \centering
  \scriptsize
  \renewcommand{\arraystretch}{1.1}
  \setlength\tabcolsep{5pt}
  \caption{ Quantitative results on CVC-ColonDB~\cite{tajbakhsh2015automated}, ETIS~\cite{silva2014toward}, and {\tt test set} (CVC-T) of EndoScene~\cite{vazquez2017benchmark} datasets. SFA~\cite{fang2019selective} results are generated using the released code.
  }\label{tab:Generalizability}
  \begin{tabular}{cr||cccccccc}
  \hline
  \rowcolor{mygray}
  &Methods & mean Dice & mean IoU  &  $F_\beta^w$ & $S_{\alpha}$&$E_\phi^{max}$ & MAE\\
  \hline
  \multirow{4}{*}{\begin{sideways}ColonDB\end{sideways}} & 
  U-Net(MICCAI'15)~\cite{ronneberger2015u}  & 0.512 & 0.444 & 0.498 & 0.712 & 0.776 & 0.061 \\
  &U-Net++(TMI'19)~\cite{zhou2018unetplus} & 0.483 & 0.410 & 0.467 & 0.691 & 0.760 & 0.064 \\
  %&ResUNet++$^\dagger$~\cite{jha2019resunetplus} & n/a  & n/a & n/a & n/a & n/a & n/a\\
  &SFA (MICCAI'19)~\cite{fang2019selective} & 0.469 & 0.347 & 0.379 & 0.634 & 0.765 & 0.094\\
  %CPD-CVPR'19~\cite{wu2019cascaded} & 0.912 & 0.877 & 0.887 & 0.805 & 0.887 & 0.882 & 0.921 & 0.961 & 0.014 \\
  %SCRNet-ICCV'19~\cite{wu2019stacked} & 0.931 & 0.818 & 0.846 & 0.783 & 0.846 & 0.831 & 0.907 & 0.960 & 0.021 \\
  %SINet-CVPR'20~\cite{fan2020Camouflage} & 0.967 & 0.818 & 0.865 & 0.821&  0.865& 0.800 & 0.921 & \textbf{0.984} & 0.017\\
  &\textbf{\ourmodel~(Ours)} & \textbf{0.709} & \textbf{0.640} & \textbf{0.696} & \textbf{0.819} & \textbf{0.869} & \textbf{0.045}\\
  \hline
  \hline
  \multirow{4}{*}{\begin{sideways}ETIS\end{sideways}} &
  U-Net (MICCAI'15)~\cite{ronneberger2015u}  & 0.398 & 0.335 & 0.366 & 0.684 & 0.740 & 0.036 \\
  &U-Net++ (TMI'19)~\cite{zhou2018unetplus}  & 0.401 & 0.344 & 0.390 & 0.683 & 0.776 & 0.035 \\
  %&ResUNet++$^\dagger$~\cite{jha2019resunetplus} & n/a  & n/a & n/a & n/a & n/a & n/a\\
  &SFA (MICCAI'19)~\cite{fang2019selective} & 0.297 & 0.217 & 0.231 & 0.557 & 0.633 & 0.109\\
  %CPD-CVPR'19~\cite{wu2019cascaded} & 0.912 & 0.877 & 0.887 & 0.805 & 0.887 & 0.882 & 0.921 & 0.961 & 0.014 \\
  %SCRNet-ICCV'19~\cite{wu2019stacked} & 0.931 & 0.818 & 0.846 & 0.783 & 0.846 & 0.831 & 0.907 & 0.960 & 0.021 \\
  %SINet-CVPR'20~\cite{fan2020Camouflage} & 0.967 & 0.818 & 0.865 & 0.821&  0.865& 0.800 & 0.921 & \textbf{0.984} & 0.017\\
  &\textbf{\ourmodel~(Ours)}  & \textbf{0.628} & \textbf{0.567} & \textbf{0.600} & \textbf{0.794}  & \textbf{0.841} & \textbf{0.031}\\
  \hline
  \hline
  \multirow{4}{*}{\begin{sideways}CVC-T\end{sideways}} &
  U-Net (MICCAI'15)~\cite{ronneberger2015u}  & 0.710 & 0.627 & 0.684 & 0.843 & 0.876 & 0.022\\
  &U-Net++ (TMI'19)~\cite{zhou2018unetplus}   & 0.707 & 0.624 & 0.687 & 0.839 & 0.898 & 0.018\\
  %&ResUNet++$^\dagger$~\cite{jha2019resunetplus} & n/a  & n/a & n/a & n/a & n/a & n/a\\
  &SFA (MICCAI'19)~\cite{fang2019selective} & 0.467 & 0.329 & 0.341 & 0.640 & 0.817 & 0.065\\
  %CPD-CVPR'19~\cite{wu2019cascaded} & 0.912 & 0.877 & 0.887 & 0.805 & 0.887 & 0.882 & 0.921 & 0.961 & 0.014 \\
  %SCRNet-ICCV'19~\cite{wu2019stacked} & 0.931 & 0.818 & 0.846 & 0.783 & 0.846 & 0.831 & 0.907 & 0.960 & 0.021 \\
  %SINet-CVPR'20~\cite{fan2020Camouflage} & 0.967 & 0.818 & 0.865 & 0.821&  0.865& 0.800 & 0.921 & \textbf{0.984} & 0.017\\
  &\textbf{\ourmodel~(Ours)}  & \textbf{0.871} & \textbf{0.797} & \textbf{0.843} & \textbf{0.925} & \textbf{0.972} & \textbf{0.010} \\
  \hline
  %\vspace{-10pt}
  \end{tabular}
\end{table*}

\begin{table*}[t!]
  \centering
  \scriptsize
  \renewcommand{\arraystretch}{1.1}
  \setlength\tabcolsep{5.0pt}
  \caption{Training and inference analysis (same platform) on CVC-ClinicDB~\cite{bernal2015wm} dataset. We record the \#epochs when the model converges. Lr = learning rate.
  }\label{tab:Params}
  \begin{tabular}{cr||ccrccc}
  \hline
  \rowcolor{mygray}
   & Methods & Epoch & Lr & Training & Inference& mean Dice \\
  %\hline
  %\multicolumn{10}{c}{Current state-of-the-art polyp segmentation results} \\
  \hline
  \multirow{4}{*}{\begin{sideways} CVC-612\end{sideways}} &
  U-Net (MICCAI'15)~\cite{ronneberger2015u} & 30 & 3e-4 & $\sim$40 minutes& $\sim$8fps & 0.823 \\
  & U-Net++ (TMI'19)~\cite{zhou2018unetplus} & 30 & 3e-4 & $\sim$45 minutes& $\sim$7fps & 0.794 \\
  %ResUNet$^\dagger$~\cite{zhang2018road}  & n/a & & n/a & n/a & 0.451 & n/a  \\
  %~~~~~~~~~~ResUNet-mod$^\dagger$~\cite{zhang2018road} & n/a & n/a & n/a & n/a & 0.779 & n/a\\
  %~~~~~~~~~~ResUNet++$^\dagger$~\cite{jha2019resunetplus}~~~~~~& n/a & n/a & n/a & n/a & 0.796 & n/a \\
  % NOTES: 150+200+150 = 500 epoch
  & SFA (MICCAI'19)~\cite{fang2019selective} & 500 & 1e-2 & $>$20 hours & $\sim$40fps & 0.700 \\
  %\textbf{\ourmodel~(Ours)} &  & & & & & & & \\
  %\hline
  %\multicolumn{10}{c}{Current state-of-the-art object segmentation baselines} \\
  %\hline
  %CPD-CVPR'19~\cite{wu2019cascaded}& 29.233 & & $\sim$180 minutes & & 0.887 & 0.921 \\
  %SCRNet-ICCV'19~\cite{wu2019stacked} & 25.226 & & $\sim$100 minutes& & 0.846 & 0.907\\
  %SINet-CVPR'20~\cite{fan2020Camouflage} & 0.967 & 0.818 & 0.865 & 0.821&  0.865& 0.800 & 0.921 & \textbf{0.984} & 0.017\\
  %\hline
  %\hline
  &\textbf{\ourmodel~(Ours)} & \textbf{20} & \textbf{1e-4} & \textbf{$\sim$30 minutes} & \textbf{$\sim$50fps}  & \textbf{0.899} \\
  \hline
  \end{tabular}
\end{table*}

\begin{figure*}[t!]
	\centering
	\begin{overpic}[width=\textwidth]{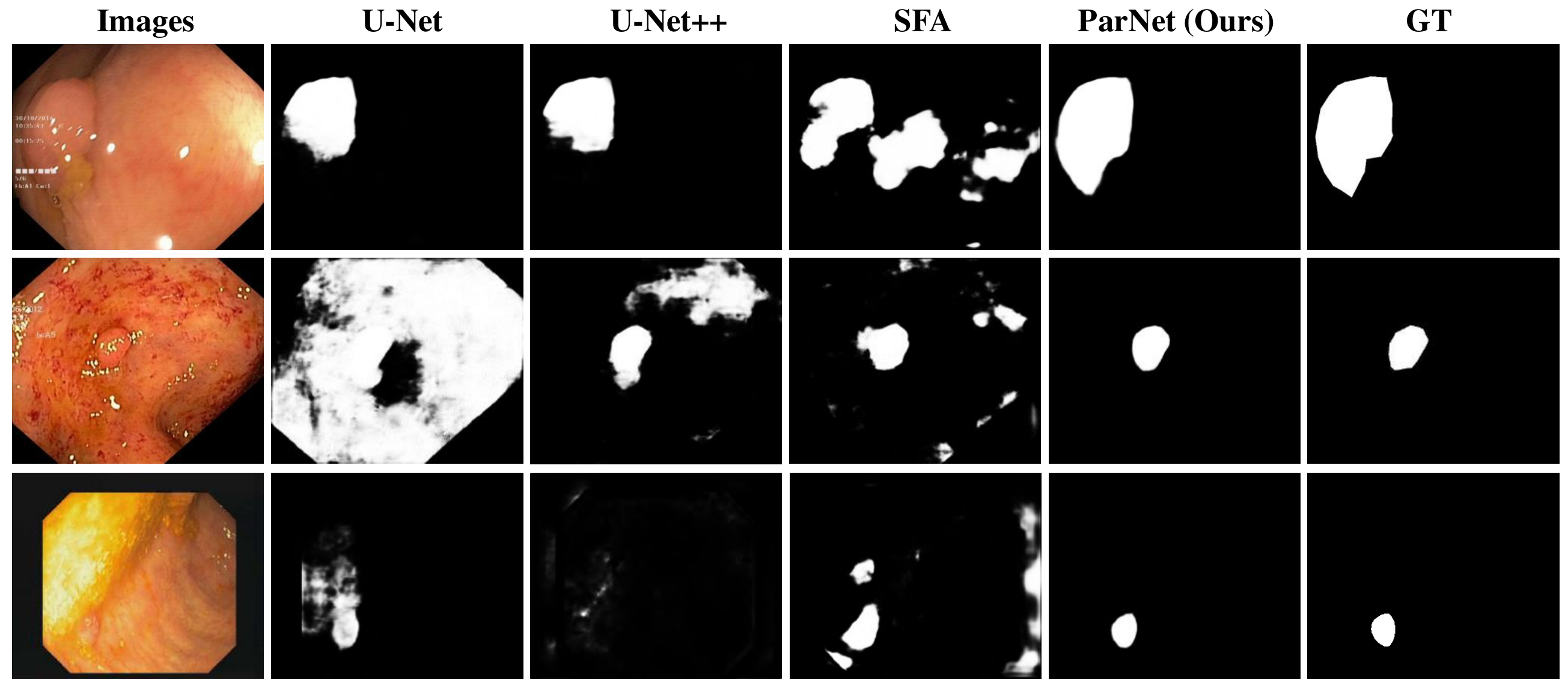}
    \end{overpic}
    %\vspace{-10pt}
	\caption{Qualitative results of different methods. 
	%Refer to the  \supp{Supp.pdf} for more details.
	}
    \label{fig:Results}
\end{figure*}

\noindent\textbf{Qualitative Results.} In \figref{fig:Results}, we provide the polyp segmentation results of our \ourmodel~on the Kvasir {\tt test set}. Our model can precisely locate and segment the polyp tissues in many challenging cases, such as varied size, homogeneous regions, different kinds of texture, \etc.\\

\noindent\textbf{Training and Inference Analysis.} In \tabref{tab:Params}, we present the training time, and inference time of \ourmodel~and current SOTA approaches. 
The running times of all compared models are tested on an Intel i9-9820X CPU and a TITAN RTX GPU with 24GB memory.
%Although our model has a larger parameter size, 
As shown, our model achieves convergence with only 20 epochs ($\sim$0.5 hours) of training. One reason is that the parallel structure of our \ourmodel~provides a short connection way to back-propagate the loss to the early layer in the decoder path (red flow of map in \figref{fig:Framework}). Moreover, the side-outputs also relieve the vanishing gradient problem and guide the early layer training. Note that our \ourmodel~runs at a real-time speed of $\sim$50fps for a 352$\times$352 input, which guarantees our method can be implemented in colonoscopy video.

%Without any pre-/post-processing, our \ourmodel~runs at real-time speed of 50fps for a 352$\times$352 input. \fdp{Owing to its simplicity, our model only requires 0.5 hours (20 epochs) can achieve strong learning and generalization ability.} \\

\subsection{Ablation Study}\label{sec:ablation}
In this section, we test each component of our \ourmodel~on the \textit{seen} and \textit{unseen} datasets to provide deeper insight into our model. \\
\noindent\textbf{Effectiveness of PPD.} 
We investigate the importance of the cascaded mechanism (parallel partial decoder, PPD). From \tabref{tab:Ablation}, we observe that No.2 (backbone + PPD) outperforms No.1 (backbone), clearly showing that the cascaded mechanism is necessary for increasing performance. Note that our PPD is only deployed on the high-level features, which greatly reduces the training time (See \tabref{tab:Params}, \textit{Inference} = $\sim$50fps) of the model. \\
\noindent\textbf{Effectiveness of RA.} We further investigate the contribution of the reverse attention. The results are listed in the first and third column of \tabref{tab:Ablation}. We observe that No.3 improves the backbone (No.1) performance on the CVC-612, increasing the mean Dice from 0.747 to 0.888 and the structure measure $S_\alpha$ from 0.735 to 0.912. These improvements suggest that introducing reverse attention component can enable our model to accurately distinguish true polyp tissues. \\
\noindent\textbf{Effectiveness of PPD \& RA.} To assess the combination of the PPD and RA modules, we test the performance of No.4 (PPD + RA + Backbone). As shown in \tabref{tab:Ablation}, our \ourmodel~(No.4) is generally better than other settings (No.1$\sim$No.3). 
In addition, \ourmodel~outperforms four SOTA models on all datasets tested, with significant improvements ($>$5\%), making it a robust, unified architecture that can help promote future research in polyp segmentation.

\begin{table}[t]
\centering
\scriptsize
\renewcommand{\arraystretch}{1.1}
\renewcommand{\tabcolsep}{3pt}
\caption{Ablation study for \ourmodel~on the CVC-612 and CVC300 datasets.
%We investigate the reverse attention (RA), and the parallel partial decoder (PPD).
}\label{tab:Ablation}
\begin{tabular}{r||ccc|ccc}
 \hline
 \rowcolor{mygray}
 \multicolumn{1}{r||}{Settings}  & \multicolumn{3}{c|}{CVC-612 (\textit{\textbf{seen}})} &\multicolumn{3}{c}{CVC300 (\textit{\textbf{unseen}})}   \\
 \rowcolor{mygray}
 &  mean Dice  &  mean IoU  &  $S_\alpha$  &  mean Dice  &  mean IoU  &  $S_\alpha$ \\
 \hline
 Backbone (No.1)            & 0.747 & 0.668 & 0.735 & 0.726 & 0.631 & 0.670 \\
 PPD + Backbone (No.2)      & 0.865 & 0.798 & 0.902 & 0.824 & 0.734 & 0.893\\
 RA + Backbone (No.3)       & 0.888 & 0.845 & 0.912 & 0.871 & \textbf{0.800} & 0.888 \\
 PPD + RA + Backbone (No.4) & \textbf{0.899} & \textbf{0.849} & \textbf{0.936} & \textbf{0.871} & 0.797 & \textbf{0.925} \\
 \hline
\end{tabular}
%\vspace{-5pt}
\end{table}

%and we argue that as many other features may play more important roles in overall survival such as histological and genetic features but unfortunately, they are not available in this challenge, a linear regression model was the safest option to minimize the test errors, although at the cost of its expressiveness.

\section{Conclusion}\label{conclusion}
We have presented a novel architecture, \ourmodel, for automatically segmenting polyps from colonoscopy images.
%\ourmodel efficiently integrates a cascaded mechanism and a reverse attention module with the parallel connection.
Extensive experiments demonstrated that \ourmodel~consistently outperforms all state-of-the-art approaches by a large margin ($>$5\%) across five challenging datasets. Furthermore, \ourmodel~achieves a very high accuracy (mean Dice = 0.898 on Kvasir dataset) without any pre-/post-processing. Another advantage is that \ourmodel~is universal and flexible, meaning that more effective modules can be added to further improve the accuracy.
%In this work, we present a simple implementation (\ourmodel) of this framework with end-to-end training.
%Finally, \ourmodel~is a simple and lightweight neural network, indicating high training efficiency.
Compared with current top-ranked SFA models, \ourmodel~can achieve strong learning, generalization ability, and real-time segmentation efficiency. %(\textbf{$\sim$50fps}). 
%Our implementation will be publicly available.
We hope this study will offer the community an opportunity to explore more powerful models on the related topics such as lung infection segmentation~\cite{fan2020inf}/classification~\cite{wu2020jcs}, or even on the upstream task, \etc.

\bibliographystyle{splncs03}
\bibliography{PolypSegmentation}

\end{document}